\documentclass[fleqn,usenatbib]{mnras}

\usepackage[T1]{fontenc}

\DeclareRobustCommand{\VAN}[3]{#2}
\let\VANthebibliography\thebibliography
\def\thebibliography{\DeclareRobustCommand{\VAN}[3]{##3}\VANthebibliography}

\usepackage{graphicx}	
\usepackage{amsmath}	
\usepackage{amssymb}	
\usepackage{lscape}
\usepackage{ae,aecompl,multirow}
\usepackage{soul}

\title[Advected Electrons in the Knots of AGN Jets]{Advection of Accelerated Electrons in Radio/X-ray Knots of AGN Jets}

\author[Amal A. Rahman et al.]{
Amal A. Rahman$^{1}$\thanks{E-mail: amalar.amal@gmail.com},
S. Sahayanathan$^{2,3}$\thanks{E-mail:sunder@barc.gov.in},
and P. A Subha $^{1}$
\\
$^{1}$Department of Physics,Farook College,  Calicut University, Kerala-673632, India\\
$^{2}$Astrophysical Sciences Division, Bhabha Atomic Research Centre, Mumbai - 400085, India\\
$^{3}$Homi Bhabha National Institute, Mumbai 400094, India\\
}

\date{Accepted 2022 June 14. Received 2022 June 12; in original form 2022 March 2}

\pubyear{2022}

\begin{document}
\label{firstpage}
\pagerange{\pageref{firstpage}--\pageref{lastpage}}
\maketitle

\begin{abstract}
The X-ray emission from the knots of the kilo-parsec scale jet of active galactic nuclei (AGN) suggests the 
high energy emission process is different from the radio/optical counterpart. Interpretation based on 
the Inverse Compton scattering of cosmic microwave photons has been ruled out through \emph{Fermi} gamma-ray 
observations for low redshift sources. As an alternate explanation, synchrotron emission from a different electron population is suggested. We propose a model considering the advected electron distribution from the sites of particle acceleration in AGN knots. This advected electron distribution is  significantly different from the accelerated electron distribution and satisfies the requirement of the second electron population. The synchrotron 
emission from the accelerated and the advected electron distribution can successfully reproduce the observed radio--to--X-ray
fluxes of the knots of 3C\,273. For the chosen combination of the model parameters, the spectrum due to inverse Compton scattering of cosmic microwave photons 
falls within the \emph{Fermi} gamma-ray upper limits.

\end{abstract}

\begin{keywords}
galaxies: active -- galaxies: individual: 3C\,273 -- galaxies: jets -- X-rays: galaxies
\end{keywords}



\section{Introduction}


The jets from active galactic nuclei (AGN) can extend up to kpc/Mpc scales with bright emission regions 
embedded called  knots \citep{2006ARA&A..44..463H}. The knots of AGN jet have been well resolved in 
radio and optical wavebands. With the advent of \emph{Chandra}, with its superior spatial 
resolution, today we know most of these knots are even bright in X-rays. The mechanism 
responsible for radio-to-optical emission from the knots is understood to be synchrotron process;
where a relativistic electron distribution loses its energy as radiation in the jet magnetic field \citep{2005ApJ...622..797K, 2010ApJ...710.1017Z}. 
However, the X-ray emission process is modeled either as synchrotron or inverse compton 
scattering of soft target photons depending up on the hardness of optical-to-X-ray spectral
index ($\alpha_{ox}$) in comparison to radio-to-optical spectral 
index ($\alpha_{ro}$) \citep{2006ApJ...648..900J}.  
When $\alpha_{ox}>\alpha_{ro}$, the X-ray emission can be interpreted as synchrotron emission
from a broken power-law electron distribution. On the other hand, when $\alpha_{ox}<\alpha_{ro}$ it 
demands convex (concave upward) electron distribution which is difficult to obtain and hence it is attributed to inverse compton
emission or emission from a different/second electron population \citep{2004ApJ...613..151A, 2004ApJ...615..161H,2005ApJ...622..797K,2006ApJ...648..910U}.

The X-ray emission from the knots, when attributed to inverse compton process;  viable source of the 
target photons can be the synchrotron photons themselves, commonly referred to as Synchrotron Self 
Compton (SSC) process.  However, this interpretation is disfavored since it demands a magnetic 
field that deviates largely from the equipartition condition \citep{2006ARA&A..44..463H, 2010ApJ...710.1017Z}. Alternatively, \cite{2000ApJ...544L..23T} showed if the 
AGN jets are relativistic even at kpc scales, then the ambient cosmic microwave background (CMB) will be 
relativistically boosted in the knot frame and can supersede the synchrotron photon energy density.
Inverse Compton scattering of this CMB (IC/CMB) can successfully explain the X-ray emission from the knots 
of PKS\,0637-752 while maintaining the equipartition condition. The IC/CMB model is also found to 
be the preferred emission mechanism to explain the X-ray knots for several other AGN jets \citep{2002ApJ...571..206S, 2006AN....327..227J, 2007ApJ...662..900T,2011ApJ...739...65P, 2012ApJ...748...81K, 2015ApJ...807...48S}.

Modeling the X-ray knots through IC/CMB emission suggests, this spectral component peaks at $\gamma$-ray 
energies. Hence, a crucial prediction by IC/CMB model is that the X-ray bright kpc scale jets of nearby AGN can 
be detectable by \emph{Fermi} space telescope operating at $\gamma$-ray energies \citep{2006ApJ...653L...5G}. 
On the contrary, $\gamma$-ray observation of 3C\,273 by \emph{Fermi} during 2008 to 2013 resulted only in
flux upper limits, even though the IC/CMB model for the knots/jet of the source predicted 
detectable flux \citep{2014ApJ...780L..27M}. A similar result was also obtained for the source PKS\,0637-752,  where
six years of \emph{Fermi} observation did not support the IC/CMB origin of the X-ray emission \citep{2015ApJ...805..154M}. 
Apart from these sources, the \emph{Fermi} non-detection of four more sources PKS\,1136-135, 
PKS\,1229-021, PKS\,1354+195 and PKS\,2209+080 with prominent X-ray jets further questioned the validity of IC/CMB 
emission as a plausible interpretation for the X-ray emission from the jets of low redshift AGN. Nevertheless, IC/CMB emission is still preferred
for the sources which are not yet disproved by the \emph{Fermi} observations \citep{2017A&A...608A..37Z, 2019ApJ...883L...2M}.

Failure of IC/CMB model to explain the \emph{Fermi} $\gamma$-ray upper limit favors the presence of second electron 
population which is responsible for the observed X-ray emission from the knots. If we assume the electrons are 
accelerated 
at the knot sites by a shock front, then the particle injection into the upstream and the downstream region will be 
assymetric due to the compression of the downstream fluid \citep{2015ApJ...806..188L}. This can develop two independent
electron populations and can explain the radio-to-X-ray emission from the knots successfully. Alternatively, electrons
can be accelerated to ultra-high energies at the sheared boundary of the jets in addition to the shock acceleration
\citep{2021MNRAS.501.6199T}. Under this scenario, the electrons accelerated at the shock front contribute 
to radio-optical emission; while the X-ray emission is produced by the electrons undergoing shear acceleration.

The drawback of the second electron population interpretation of the X-ray knots is,
it demands ultra high energy electrons which are subject to fast cooling \citep{2020ApJ...893...41W}.
Hence, these electrons would expend all their energy close to their production site  itself 
and may not explain the extended X-ray emission from the knots. \cite{2020ApJ...893...41W}, therefore,
suggested a lepto-hadronic origin for the radio-to-X-ray emission for the knots. The radio-to-X-ray emission
from the knot can also be explained by models involving emission from hadrons alone \citep{2014MNRAS.444L..16K}.
Advantage of models involving the X-ray emission from hadrons is, they lose their energy much slower than the 
leptons and therefore can diffuse over long distances. 

In this paper, we present an emission scenario for the knots where the advection of electrons from the main 
acceleration site is considered. This can produce two distinct electron populations with a high energy 
electron distribution arising from the acceleration site and a low energy one surrounding it. 
The synchrotron emission from such combined electron
distribution is capable to explain the radio-to-X-ray emission from the knots of the jet of 3C\,273. In the 
next section, we deduce the advected electron population and explain the total  synchrotron spectrum. In section 3, we apply this model on the knots of 3C\,273 and discuss the results.
Throughout this work we consider a cosmology where 
$H_0 =71$ km s$^{-1}$ Mpc$^{-1}$, $\Omega_m = 0.27$ and $\Omega_\Lambda = 0.73$.

\section{The Advected Electron Population}
We consider a model in which the particle acceleration in the AGN knot is confined within a spherical region of size $R_0$.
The electron distribution in this region is assumed to be a broken power-law described by
\begin{align}\label{eq:bpl}
  n_0(\gamma)d\gamma=\begin{cases}
               K\gamma^{-p}d\gamma  \hspace{15mm}    \gamma_{\rm min} < \gamma < \gamma_{b}\\
              K\gamma_b^{q-p}\gamma^{-q}d\gamma  \hspace{10mm}  \gamma_{b} < \gamma < \gamma_{\rm max} 
            \end{cases}
\end{align}
Such an electron distribution can be an outcome of multiple acceleration sites embedded 
within $R_0$ \citep{2008MNRAS.388L..49S,1994PASA...11..175P}. The accelerated electrons are advected outside $R_0$ where they do not undergo further  acceleration  
but lose their energy through radiative processes and adiabatic cooling. We consider the magnetic field at the regions $R<R0$ and $R>R0$ as $B_{in}$
and $B_{out}$ respectively. 
The evolution of the electron distribution in 
the region $R>R_0$ can be described by  \citep{1962SvA.....6..317K,1997astro.ph.11129A}
\begin{equation}\label{eq:kin1}
	\frac{\partial n(\gamma,t)}{\partial t}=\frac{\partial}{\partial\gamma}\left[P(\gamma)\,n(\gamma,t)\right] - \frac{n(\gamma,t)}{t_{\rm esc}(R)} + q(\gamma)
\end{equation}
Here, $P(\gamma)$ is total energy loss rate due to radiative processes and adiabatic expansion, $t_{\rm esc}(R)$ \footnote { We express the advection term 
in terms of $t_{esc}$ as follows: for a constant velocity v, the advection can be expressed as
$-v \frac{\partial n}{\partial R} $.
Changing variable $R/v = t$ we get,
\begin{equation}\nonumber
-\frac{\partial n}{\partial t} \approx -\frac{n}{t_{esc}(R)}
\end{equation}}
\citep{1964ocr..book.....G} 
is electron escape timescale at a distance $R$ 
from $R_0$ and $q(\gamma)$ is the rate of electron injection. 

For a constant fluid velocity, the distance $R$ will be linearly 
dependent on time $t$ and this let us to express
equation (\ref{eq:kin1}) in terms of $R$ as 
\begin{equation}\label{eq:kin2}
	\frac{\partial \bar{n}(\gamma,R,R_0)}{\partial R}=\frac{\partial}{\partial\gamma}\left[\bar{P}(\gamma)\,\bar{n}(\gamma,R,R_0)\right] - \frac{\bar{n}(\gamma,R,R_0)}{R_*(R)} + n_0(\gamma)\,\delta(R-R_0)
\end{equation}
where, $R_*$ is the characteristic particle escape radius corresponding to $t_{\rm esc}$ and  
\begin{align}\label{eq:loss}
\bar{P}(\gamma)&= \frac{d\gamma}{dR}\nonumber \\ &= \xi \gamma^2 + \frac{\gamma}{R}
\end{align}
where, $\xi$ will be a function of magnetic field and the velocity of the fluid \citep{1986rpa..book.....R,1997astro.ph.11129A}.
The last term in equation (\ref{eq:kin2}) implies the particle injection happens only at $R_0$. 

The solution of equation (\ref{eq:kin2}) can be obtained through the the Green's function for $R>R_0$ as \citep{1997astro.ph.11129A}
\begin{align}\label{eq:nbar}
	\bar{n}(\gamma,R,R_0)=\frac{R_0}{R}\frac{\Gamma_0^2}{\gamma^2}\,n_0(\Gamma_0)\,{\rm exp} \left[-\int_{R_0}^{R}\frac{dx}{R_*(x)}\right]
\end{align}
where, $\Gamma_0$ defines the energy of the electron at injection radius $R_0$ which reduces to $\gamma$ at $R$ due to 
radiative and adiabatic losses. From equation (\ref{eq:loss}) we find
\begin{align}
\Gamma_0(\gamma, R)= \frac{\gamma\frac{R}{R_0}}{1-\xi\gamma R \ln \frac{R}{R_0}}
\end{align}
The maximum distance $R_{\rm max}$ up to where electrons with energy $\gamma$ will be available and can be obtained by numerically solving the transcendental equation
\begin{align}
\frac{R_{max}}{R_0}=\frac{\gamma_{max}}{\gamma}\left[1-\xi\gamma R_{max} \ln \left(\frac{R_{max}}{R_0}\right)\right]
\end{align}
and
\begin{align}
	R_{\rm max}(\gamma) \leq R_0 + \frac{1}{\xi}\left(\frac{1}{\gamma}-\frac{1}{\gamma_{\rm max}}\right)
\end{align}
where, the inequality is obtained by considering only radiative losses.
We assume the escape radius as
\begin{align}
	R_*(R) = \zeta R^\alpha
\end{align}
where, $\zeta$ and $\alpha$ are constants. 
For $\alpha = 1$, the advected electron distribution
in the region external to $R_0$ will be
\begin{align}\label{eq:diff}
	\bar{n}(\gamma, R, R_0) = \frac{R_0}{R}\frac{(\Gamma_0)^2}{(\gamma)^2}n_0(\Gamma_{0})\left(\frac{R_0}{R}\right)^{{1}/{\zeta}}
\end{align}
In Fig. 1 we show the distribution of $\bar{n}$ for different $\frac{R}{R_0}$.
The general solution for $\alpha \ne 1$ is given in Appendix \ref{app:alpha}. 

\begin{figure}
		\centering
		\includegraphics[scale=0.7, angle=0]{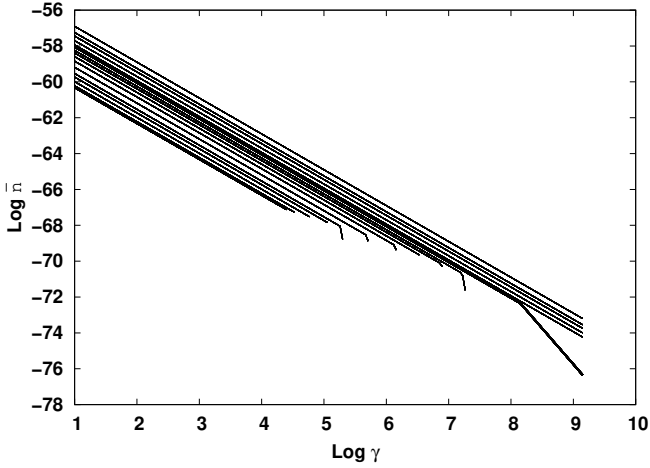}
		\caption {The diffused electron distribution at different $R>R_0$ for the cases of $\frac{R}{R_0}= 0.1, 0.2, 0.3, 0.5, 0.8, 1, 1.5, 2, 3, 8, 10, 20, 30,50, 80$ and $100$. The parameters chosen are of knot A. The bold line corresponds to $\frac{R}{R_0}= 1$}
		\label{Fig1}
\end{figure} 

\subsection{Synchrotron Spectrum}
The total number of electrons with energy $\gamma$ from the regions $R<R_0$ and $R>R_0$ can be obtained 
from equations (\ref{eq:bpl}) and (\ref{eq:diff}) as
\begin{align}
	N_{\rm tot}(\gamma)= \frac{4}{3}\pi R_0^{3} n_0(\gamma)+ 4\pi\int_{R_0}^{ {\rm MIN}(R_{\rm max},R_{\rm size})}\bar{n}(\gamma,R,R_0)R^2dR
\end{align}
where, $R_{\rm size}$ is the size of the knot. In Fig. 2,  we plot the distribution of the advected electron population for different choices of $\zeta$. The total electron distribution from the acceleration and advected region is shown in Fig. 4.
The synchrotron emissivity at frequency $\nu$ 
can be obtained by convolving the total electron distribution with the single particle emissivity $j_{\rm syn}(\gamma,\nu)$ \citep{1986rpa..book.....R}
\begin{align}
	J_{\rm syn}(\nu)= \frac{1}{4\pi}\int_{\gamma_{\rm min}}^{\gamma_{\rm max}}j_{\rm syn}(\gamma,\nu)N_{\rm tot}(\gamma)d\gamma     
\end{align}         

The observed flux on earth after accounting for the relativistic effects of jet and cosmological corrections will be 
\begin{align}\label{eq:obs_flux}
	F_{\rm obs}(\nu_{\rm obs})= \frac{\delta_D^3(1+z)}{d_L^2}J_{\rm syn}\left(\frac{1+z}{\delta_D}\nu_{\rm obs}\right)
\end{align}
Here, $d_L$ is the luminuousity distance and $\delta_D$ is the Doppler factor given by
\begin{align}
	\delta_D=\frac{1}{\Gamma(1-\beta {\rm cos}\theta)}
\end{align}
with $\Gamma$ being the bulk Lorentz factor, $\beta\,c$ the velocity and $\theta$ the viewing angle
of the relativistic jet.

\section{Results and Discussion}
The model described above, incorporating the advection of electrons from the sites of particle 
acceleration, is used to reproduce the radio/optical/X-ray fluxes from the knots of 3C\,273. The primary accelerated 
electron distribution is assumed to be a broken power-law and confined within a region $R<R_0$ with magnetic field $B_{in}$ (equation(\ref {eq:bpl})). 
The synchrotron emission by this electron distribution is used to model the X-ray flux from the knots of 3C\,273.
The advected electron distribution from the region $R>R_0$ creates a low energy excess and the synchrotron emission from this electron
population can reproduce the observed radio fluxes. 

In Fig. 3, we show the observed radio--optical--X-ray spectrum from the knots A, B1, B2, B3, C1, C2, D1 and D2H3 
along with the model curves. The dot-dash line is the synchrotron emission from the accelerated electron
distribution while the dotted line is from the advected electron population. The total emission is shown as 
solid line. The parameters used to reproduce the model curves are given in Table. 1. Besides these parameters,
we have assumed $\alpha = 1$, $\gamma_{\rm min} = 10$, $\gamma_{\rm max} = 10\gamma_b$ and $\theta=20$ degrees.

The choice of model parameters also decide whether the IC scattering of CMB photons by the total electron distribution
will exceed the \emph{Fermi} upper limits. To verify this we estimate IC/CMB spectral component using approximate analytical 
solution for a monochromatic external photon field \citep{2018RAA....18...35S}. The IC/CMB emissivity at frequency $\nu$
can be expressed as 
\begin{align}
	j_{\rm ec}(\nu) \approx \frac{c\sigma_{T}U_*}{8\pi{\nu}_*} \sqrt[]{\frac{\Gamma\nu (1+\mu)}{{\nu}_*}} N_{\rm tot}\left[\sqrt{\frac{\nu}{\Gamma {{\nu}_*}(1+\mu)}}\right]
\end{align}
Here, ${\nu}_*$ and $U_*$ are the frequency and the energy density of the external photon field and $\mu$ 
is the jet viewing angle measured in the proper frame of the AGN. In Fig. 3, the specral energy distribution due to
IC/CMB emission 
shown as dashed line. For the choice of parameters given in Table. 1, the spectrum due to IC/CMB 
falls below the \emph{Fermi} upper limits.

\begin{figure}
		\centering
		\includegraphics[scale=0.7, angle=0]{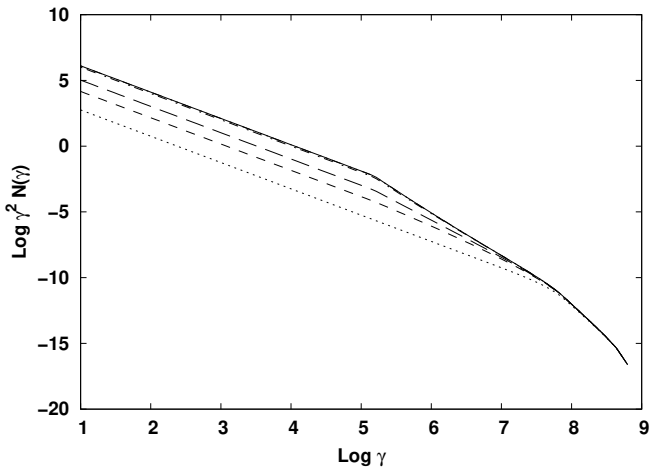}
		\caption{Advected electron population for different $\zeta$ in the case of knot A. Solid line corresponds to $\zeta$=100  and dotted line to $\zeta$=0.1. All other plots below solid line corresponds to $\zeta$=50, 10, 1  \&  0.5.}
		\label{Fig2}
\end{figure} 

\begin{figure*}
\includegraphics[angle=0,scale=0.55]{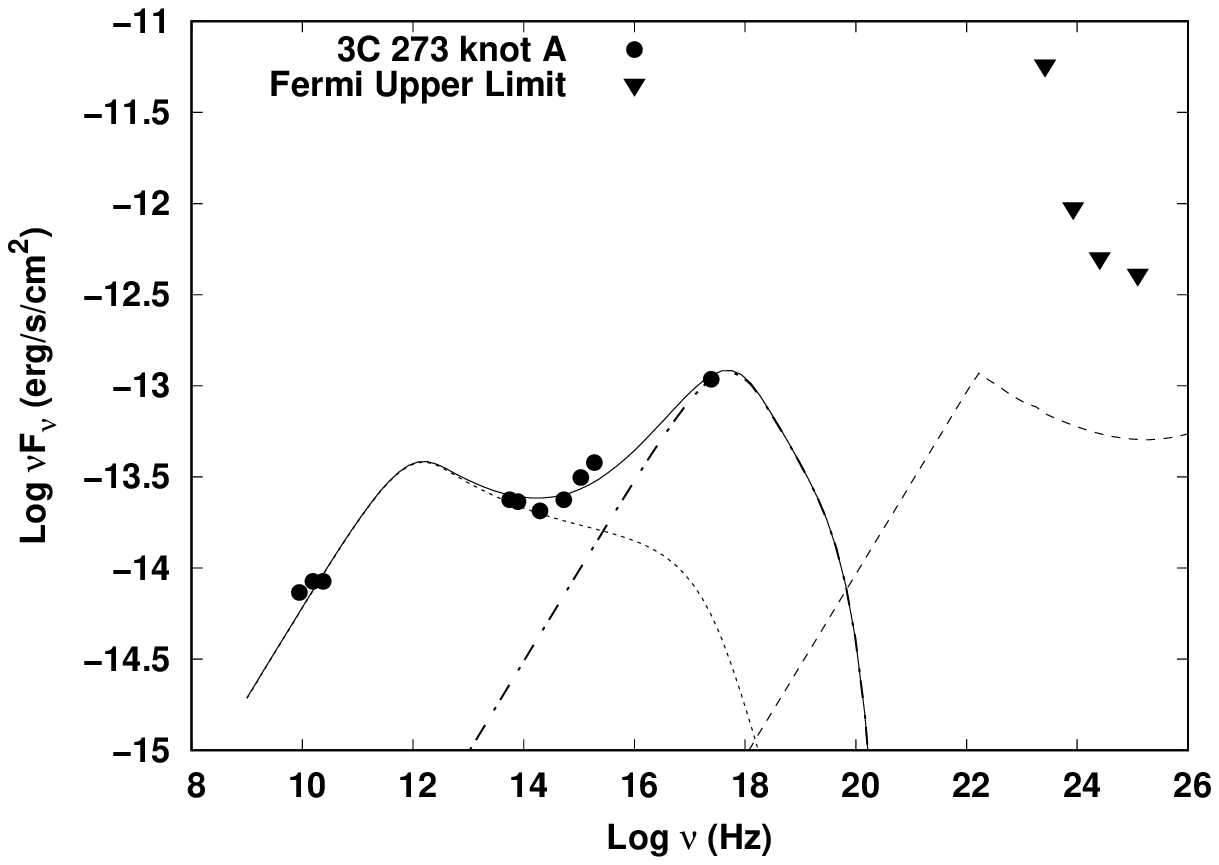}
\includegraphics[angle=0,scale=0.55]{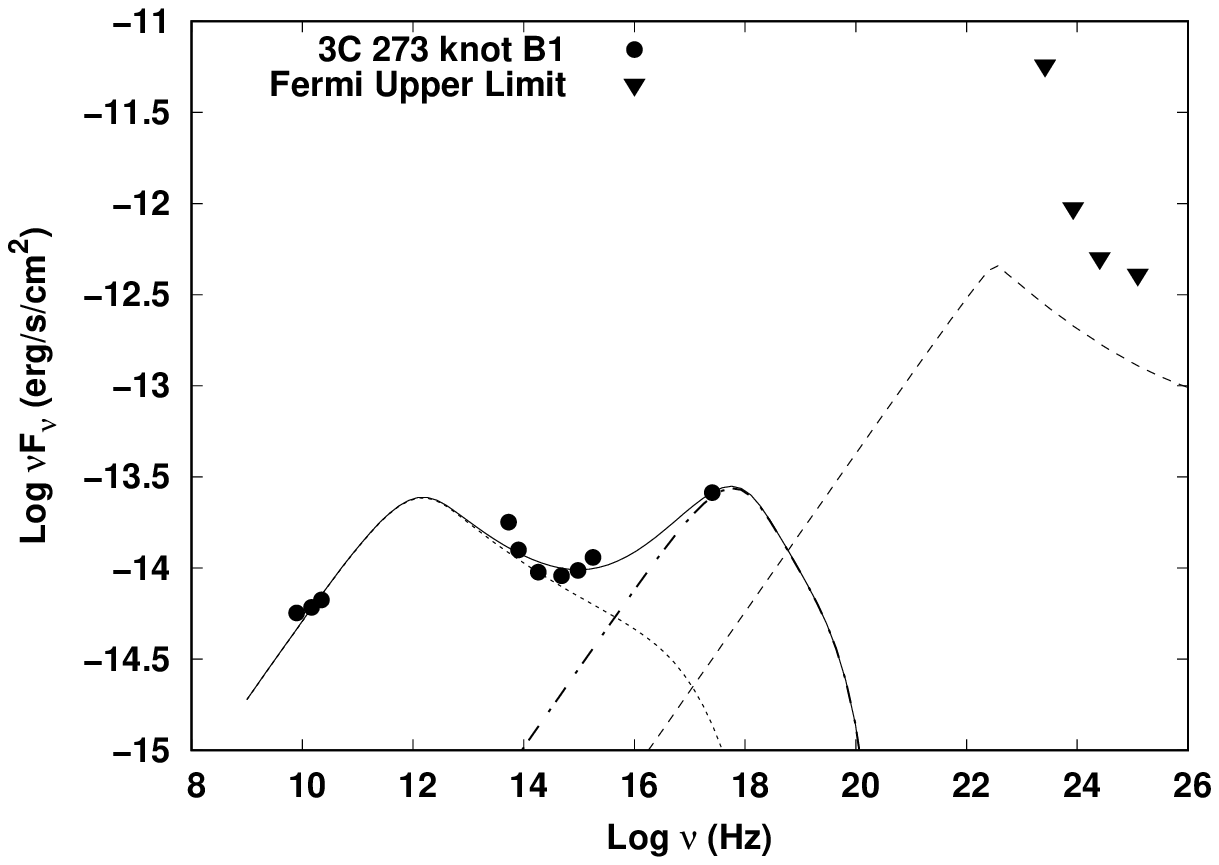}
\includegraphics[angle=0,scale=0.55]{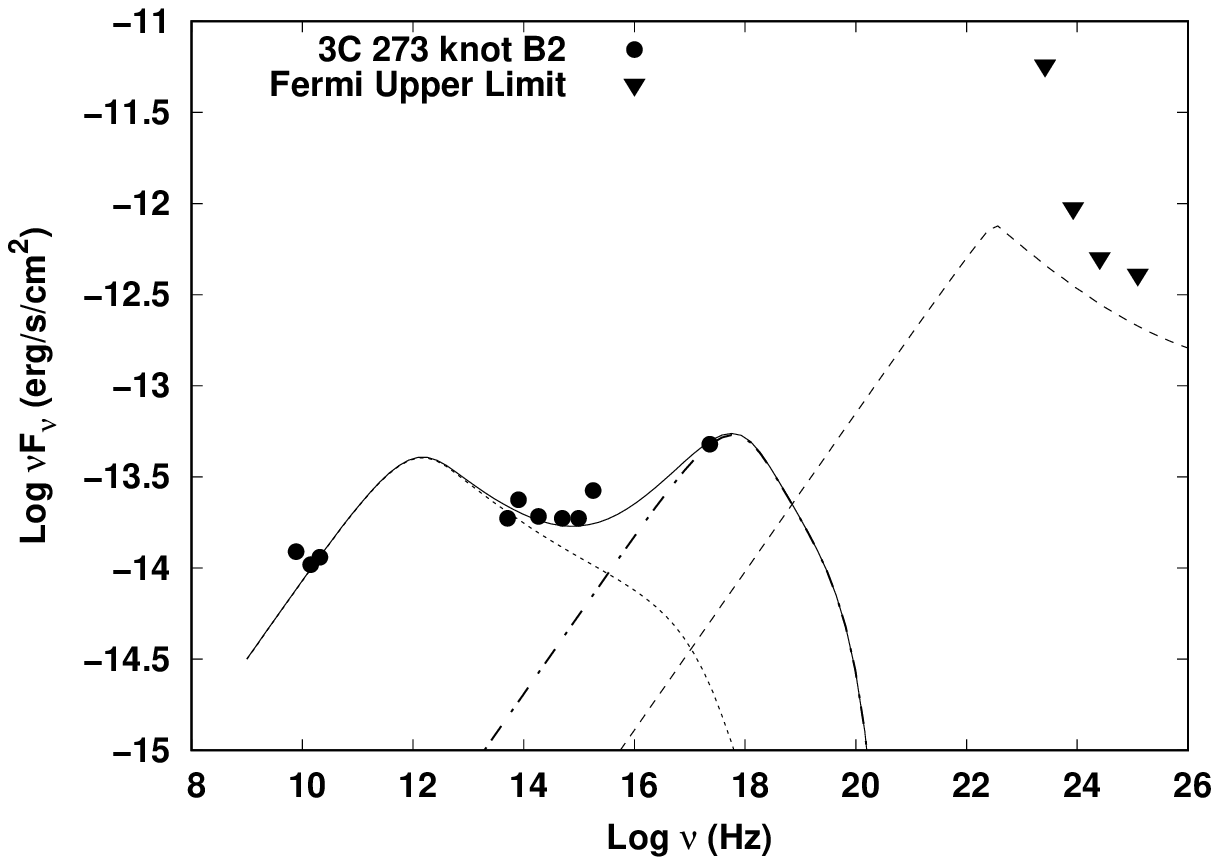}
\includegraphics[angle=0,scale=0.55]{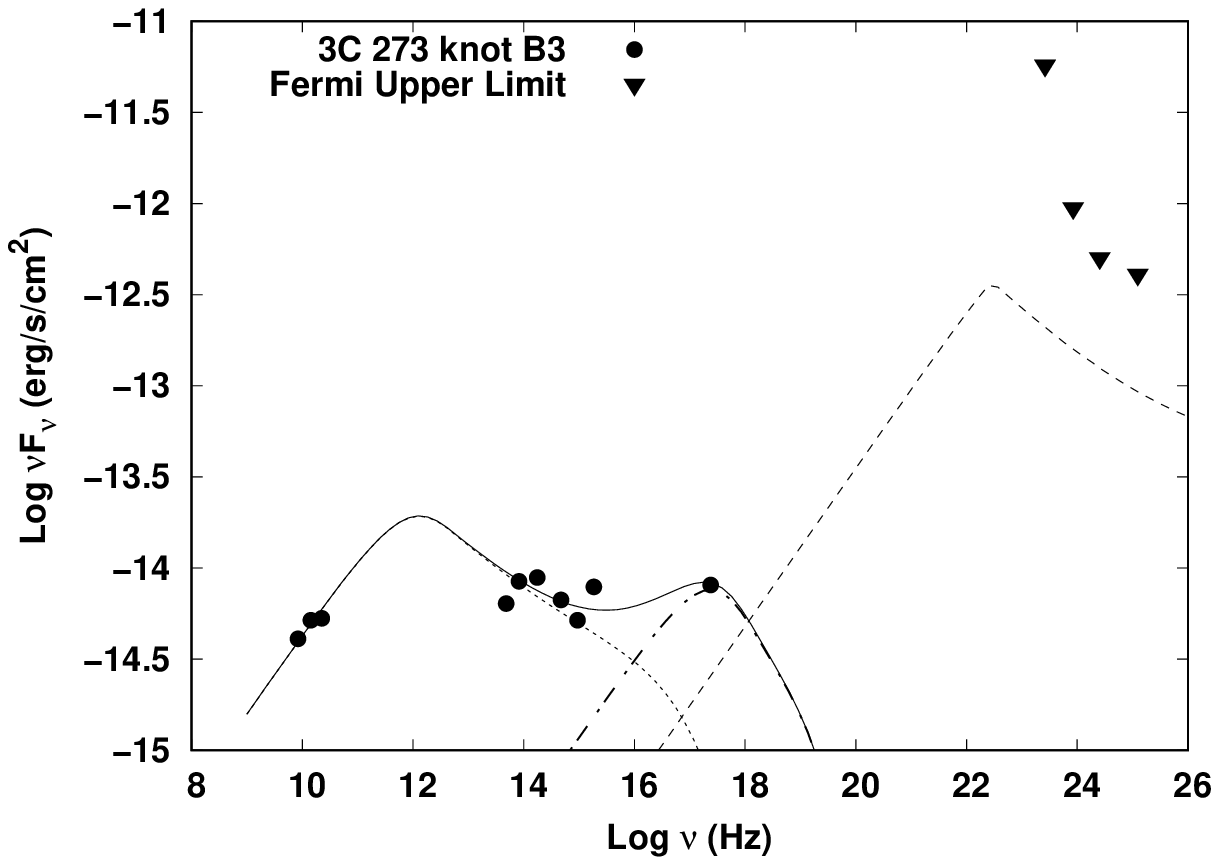}
\includegraphics[angle=0,scale=0.55]{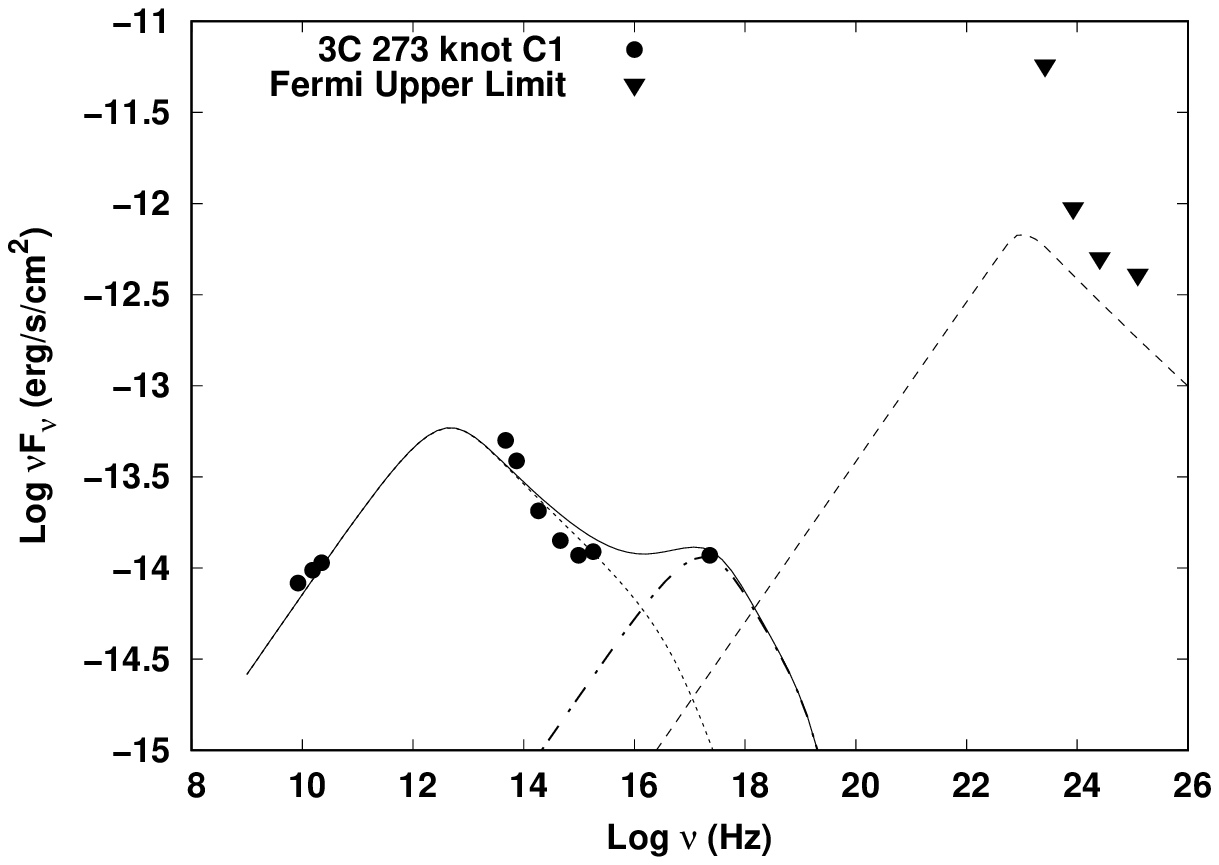}
\includegraphics[angle=0,scale=0.55]{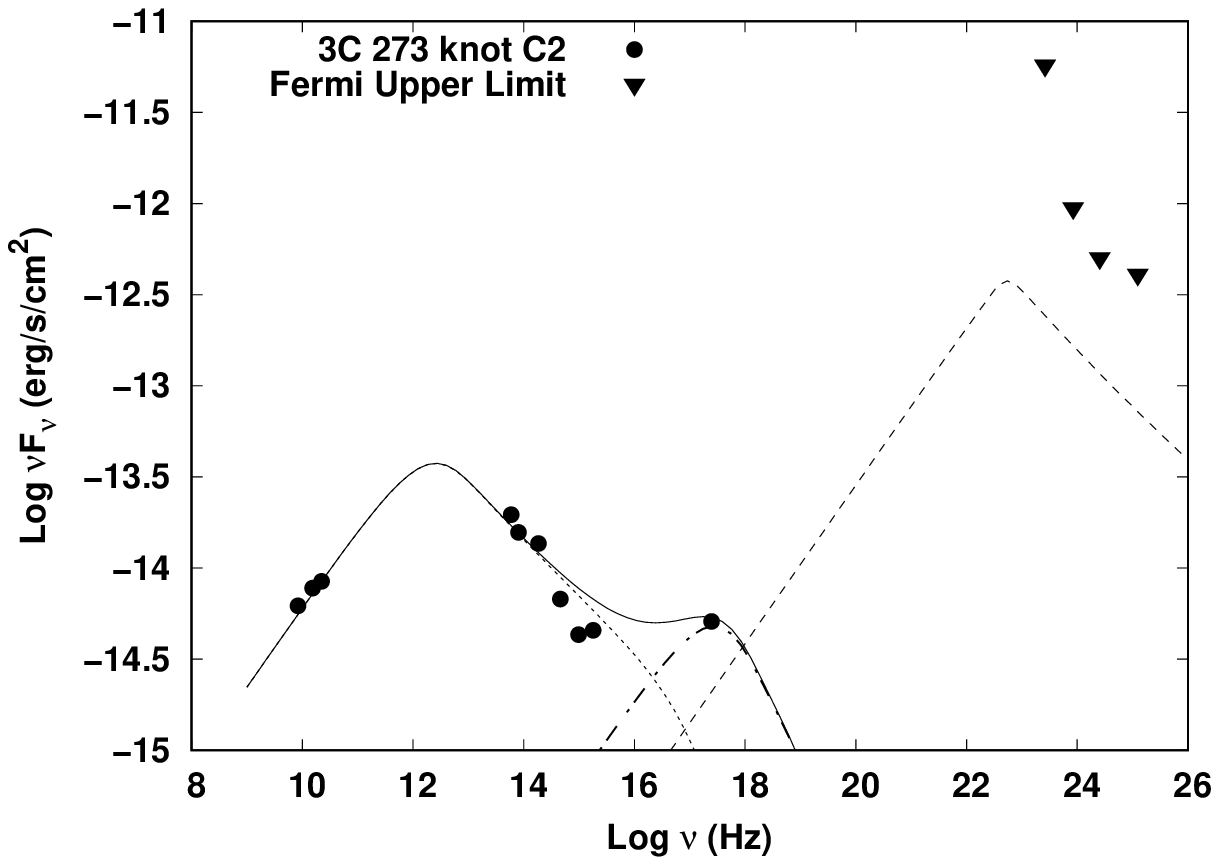}
\includegraphics[angle=0,scale=0.55]{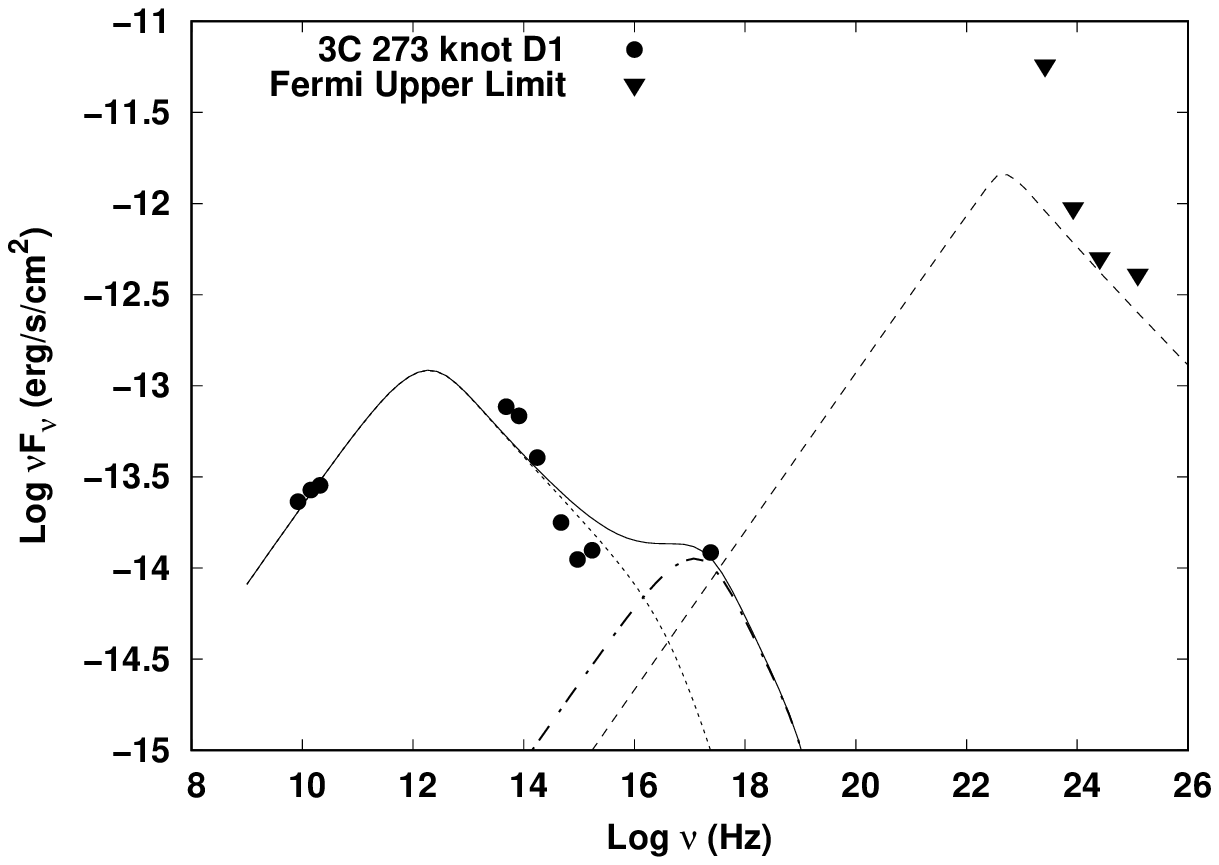}
\includegraphics[angle=0,scale=0.55]{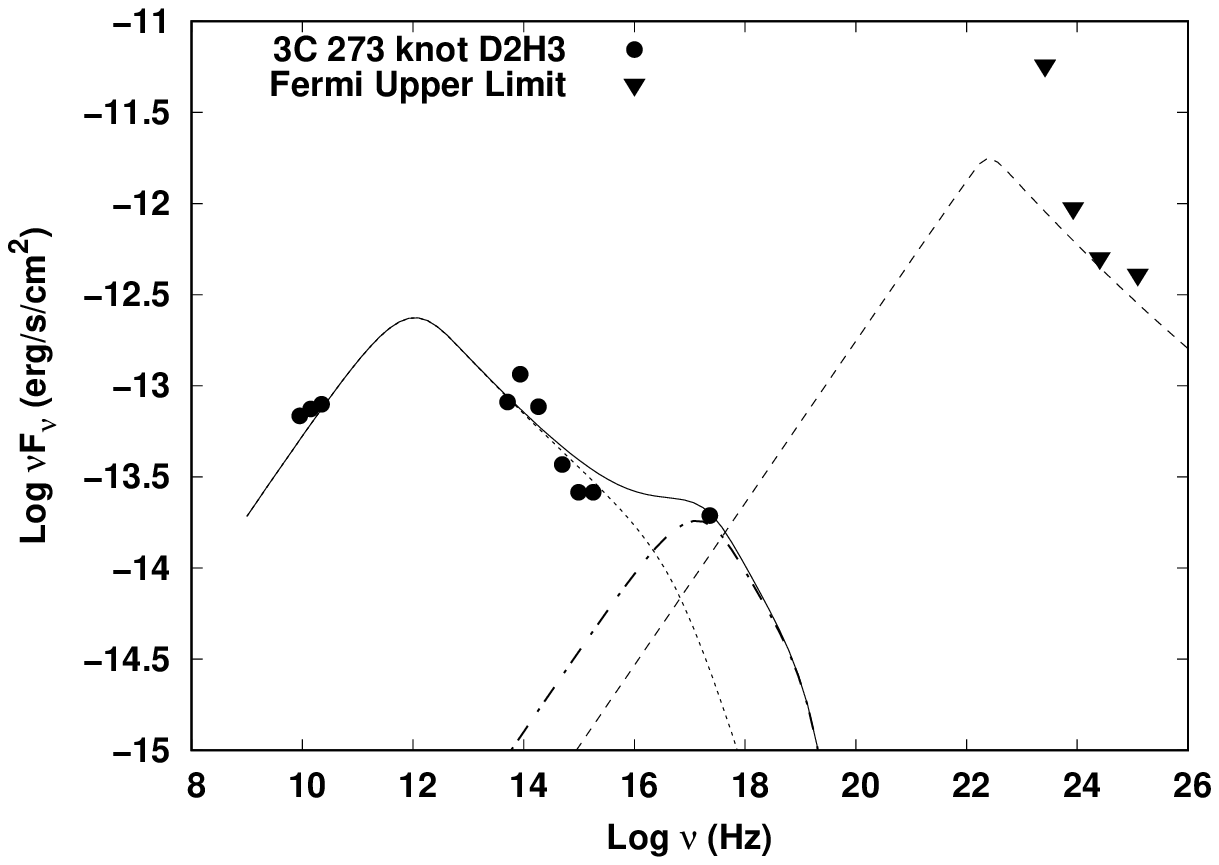}
\caption{Observed broadband SEDs (solid circles) of the knots of 3C\,273 together with the model curves. Dot-dash line represents the synchrotron emission from accelerated electron population, dotted line represents synhrotron emission from advected electron population and solid line represent total synchrotron emission. The dashed line corresponds to IC/CMB spectrum and \emph{Fermi} upper limits are denoted by inverted triangles.}\label{single1}
\end{figure*}

\begin{table*}
  \centering
\begin{tabular}{|c|c|c|c|c|c|c|c|c|c|c|}
\hline
\multirow {1}{*} Knot & $R_0(kpc)$  & $R_{size}(kpc)$     & $B_{in}(10^{-5}G)$  &$\omega = \frac{B_{in}}{B_{out}}$		&$\gamma_b (10^7)$  & $v_{ad}(10^{-2}c)$  & $p$     & $q$  & $\Gamma$ & $\zeta$   \\ \hline
A & ${0.12}$ & ${2.9}$ & ${1.5}$  &$2.08$ & ${7.5}$ & ${2.0}$ &  ${2.0}$  &  ${4.0}$    & ${1.3}$   & ${7.0}$\\
B1 & ${0.10}$ & ${5.5}$ & ${1.8}$ &${3.3}$ & ${6.5}$ & ${1.6}$ &  ${2.13}$    &  ${4.0}$  &  ${2.0}$ &  ${7.0}$ \\
B2 & ${0.10}$ & ${5.50}$ & ${2.0}$ &${3.6}$ & ${6.18}$ & ${1.6}$ & ${2.13}$ &  ${4.0}$  &  ${2.0}$ &  ${7.0}$     \\
B3 & ${0.09}$ & ${5.5}$ & ${1.3}$ &${2.4}$ & ${4.85}$ & ${1.6}$ &  ${2.13}$ &  ${4.0}$ &  ${2.0}$ &  ${7.0}$     \\
C1 & ${0.08}$ & ${8.0}$ & ${1.9}$ &${3.45}$ & ${3.64}$ & ${9.0}$ &  ${2.12}$ &  ${4.0}$  & ${1.7}$ & ${5.5}$    \\
C2 & ${0.05}$ & ${6.0}$ & ${1.9}$ &${3.17}$ & ${4.70}$ & ${5.0}$ &  ${2.13}$ &  ${4.2}$ &${1.7}$ &${9.5}$     \\
D1 & ${0.05}$ & ${6.0}$ & ${1.5}$ &$3.0$ & ${3.58}$ &   ${4.0}$ &  ${2.13}$   &  ${4.2}$ & ${1.6}$ & ${9.5}$  \\
D2H3 & ${0.05}$ & ${6.5}$ & ${1.1}$ &$2.4$ & ${4.62}$ & ${2.5}$ &  ${2.11}$ &  ${4.0}$ &  ${1.3}$  &  ${9.5}$  \\

\hline
\end{tabular}\caption{Fit parameters of radio-optical-X-ray spectrum. $R_0$ is the size of inner region; $R_{size}$ is the size of the knot; $B_{in}$ and $B_{out}$ represent the magnetic field strength at $R<R_0$ and $R>R_0$ respectively. The quantities $\gamma_b$, $v_{ad}$, p, q, $\Gamma$ and $\zeta$ represent the break energy, advection velocity, power law indices of broken power law distribution of particles, bulk Lorentz factor and the constant defining escape radius respectively.}  

\label{tab1}
\end{table*}

In this work, we have assumed the accelerated electron distribution to be a broken power-law and this can be 
an outcome of multiple acceleration processes \citep{2008MNRAS.388L..49S}. For instance, if we consider the case where the particle acceleration
happens due to a standing shock buried in a turbulent plasma, then the electrons injected into the shock front 
are already accelerated under stochastic process. If the confinement time 
at the turbulent region is longer than the region in the vicinity of the shock front, the electron distribution
accelerated through stochastic process will be harder \citep{2007Ap&SS.309..119R}. 
This eventually give rise to a broken power-law distribution
with indices governed by the ratio of acceleration to escape timescales in both the regions.

\begin{figure}
		\centering
		\includegraphics[scale=0.7, angle=0]{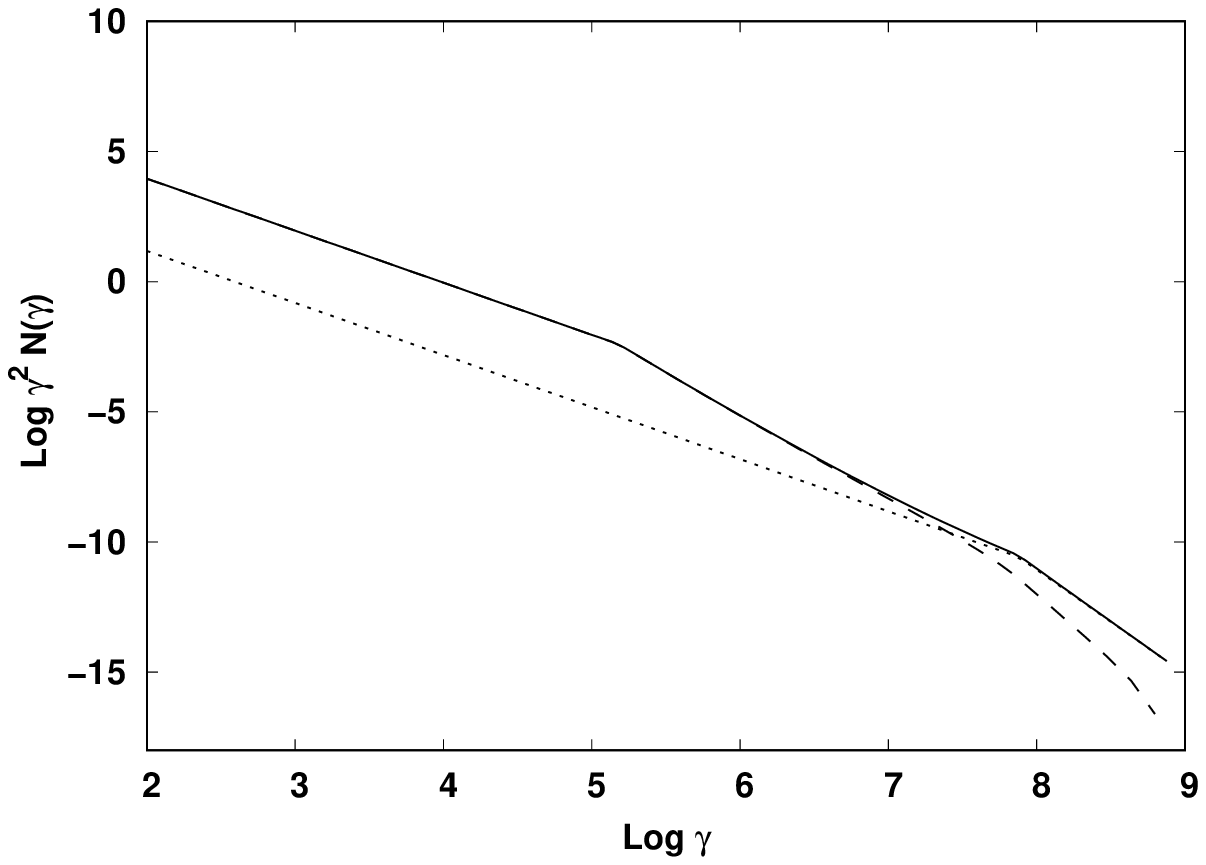}
		\caption{{{The electron distribution for the knot A of 3C\,273 . Dotted line represent accelerated  particle distribution and dashed line the advected electron distribution. Solid line is the total particle distribution.}}}
		\label{Fig1}
\end{figure} 
The model considered here demands large electron energies to explain the X-ray emission. Such large electron energies can be achieved under moderate loss or escape time scales. If we assume the escape timescale is governed by gyro radius ($R_g$), then for an electron with Lorentz factor $10^8$, $R_g$ is $10^{-3} $pc, which is much smaller than the size of inner region $R_0$. Similarly, the acceleration timescale for the given gyro-radius is approximately $R_g/c$ and is of the order of $10^5$ s \citep{1996ApJ...463..555I, 2000ApJ...536..299K}. Whereas, the synchrotron radiative loss time scale for electrons of energy $\approx 10^8$ is of orders of $10^9$s. Hence, the acceleration of electrons to such high energies is viable
under the scenario considered here.

It is interesting to study the radial evolution of the advected electron distribution which is governed by the synchrotron and adiabatic cooling(Fig.1). The integrated distribution over $R$ in combination with the accelerated electron distribution will then reflect a convex (concave upward) shape which can explain the radio--optical--X-ray spectrum of AGN knots (Fig. 4). The radial evolution of the advected electron distribution also depends up on the choice of $\alpha$. In the present study, $\alpha$ is chosen to be unity for simplicity; however, the observed knot spectrum can also be explained with $\alpha \ne 1$ and a modified set of parameters.

The synchrotron cooling time scale estimated for the radio, optical and X-ray emitting electron (for the case of $B\approx 10^{-5}$) is approximately $10^{14}$s, $10^{12}$s, and $10^{10}$s respectively\citep{1997astro.ph.11129A}. Equivalent length scale can be estimated as 100 kpc (radio), 1 kpc (optical) and  10 pc (X-ray) for a velocity $v \approx 0.1c$. This suggests the radio emitting electrons will travel the extended jet before losing its energy as compared to the X-ray emitting electrons. This is consistent with the morphological feature of the source where the radio jet extends over large scales whereas the X-ray emission dies out beyond the knot locations \citep{2006ARA&A..44..463H}. 

\emph{Fermi} non-detection of gamma-ray emission from the knots of AGN jets favored the two-electron population 
hypothesis for the radio--to--X-ray emission.
Nevertheless, IC/CMB model for the X-ray emission is still preferred for the sources 
which are not yet ruled out by \emph{Fermi} 
observations \citep{2021arXiv211108632I,2020MNRAS.497..988W,2012ApJ...748...81K}. 
Irrespective of \emph{Fermi} observations, the IC/CMB model for the X-ray emission fail to explain many  
morphological features of the AGN jet. For example, the electron energies required to 
produce the X-ray emission through IC/CMB 
process is similar to the ones emitting radio through synchrotron process. Hence, the radio/X-ray jet morphology should be comparable \citep{2006ARA&A..44..463H}. On the contrary, positional offset has been detected between the radio and X-ray maxima 
for the knots of many AGN \citep{2003ApJ...593..169H, bai2003radio}.
Similarly, IC/CMB model demands significant jet speed to explain the X-ray emission which in turn predict 
one-sided jets due to relativistic debeaming of the counter jet. However, the detection of faint counterjet
in Pictor A disfavors the IC/CMB origin of the X-ray emission \citep{2016MNRAS.455.3526H}.

\section{Summary}
The explanation for the radio--to--X-ray flux from the knots of many AGN jets through synchrotron process 
suggests the underlying electron distribution to be convex (concave upward). Production of such a 
distribution demands an acceleration process that can provide excessive power at high electron energies.
Alternatively, the synchrotron emissions from two spatially separated electron populations are
capable to explain the radio--to--X-ray emission from the AGN knots. 
In this work, we show such distributions can naturally 
arise if we incorporate the advection of accelerated electrons from the sites of particle acceleration. 
The synchrotron emission
received from the combination of accelerated and advected electron distributions are used to model the radio/optical/X-ray 
fluxes from the knots of 3C\,273.  The model parameters are chosen to reproduce the observed fluxes while the predicted IC/CMB
emission fall within the \emph{Fermi} upper limits. 

The model assumes the accelerated electron distribution to occupy a spherical region and the advected electrons
in a spherical shell around it. If we relax this spherical symmetry, the proposed model is also capable to 
explain the observed offsets between the radio, optical and X-ray knot positions.
Reproduction of the radio, optical and X-ray observations of the knots of 3C\,273 using this model suggests the
spectral energy distribution of these knots to peak at infrared frequencies. High resolution observation at 
this frequency in future have the potential to constrain/rule out the model presented in this work.

\section*{Acknowledgements}

We gratefully acknowledge the anonymous referee for very constructive comments. A.A.R also acknowledge the financial support provided by University Grants Commission, Govt. of India.

\section*{Data Availability}

The data and the codes used in this work will be shared on the
reasonable request to the corresponding author Amal A. Rahman (email:amalar.amal@gmail.com)



\bibliographystyle{mnras}
\bibliography{ref} 




\appendix 

\section{Particle distribution for the case $ \alpha \ne 1$ }
\label{app:alpha}
With definitions of $R_*$ as in section 2,  solution to equation (\ref{eq:nbar})  for the case of $\alpha \ne 1$ is given by

\begin{align}\label{eq:alpha}
n(\gamma, R, R_0) = n_0 (\Gamma_0) \frac{\Gamma_{0}^{2}} {\gamma^2} \exp\left[\frac{-1}{\zeta(-\alpha+1)}\left({R^{-{\alpha}+1}-R_0^{-{\alpha}+1}}\right)\right]
\end{align}



\bsp	
\label{lastpage}

 \end{document}